\documentclass[13pt]{article}
\usepackage[dvips]{graphicx}
\begin{document}
\begin{center}
{\LARGE\bf Dynamics of Tachyon and Phantom Field beyond the
Inverse Square Potentials}

 \vskip 0.15 in
 $^\dag$Wei Fang$^{1, 2}$ and Hui-Qing Lu$^{2, 3}$\\
\small {$^1$\cal $ Department ~of ~Physics, ~Shanghai~ Normal~ University, ~Shanghai, ~200234, ~P.R.China$\\
$^2$\cal$ The ~Shanghai~ Key ~Lab ~for ~Astrophysics, ~Shanghai,
~200234, ~P.R.China$\\
 $^3$\cal$Department ~of ~Physics, ~Shanghai~ University,
~Shanghai, ~200444, ~P.R.China$\\}
 \footnotetext{$\dag$ \ \ wfang@shnu.edu.cn}

 \vskip 0.5
in  \centerline{\bf Abstract} \vskip 0.2 in
\begin{minipage} {5.8in} {\hspace*{10pt}\small

We investigate the cosmological evolution of the tachyon and
phantom-tachyon scalar field by considering the potential
parameter $\Gamma$($=\frac{V V''}{V'^2}$) as a function of another
potential parameter $\lambda$($=\frac{V'}{\kappa V^{3/2}}$), which
correspondingly extends the analysis of the evolution of our
universe from two-dimensional autonomous dynamical system to the
three-dimension. It allows us to investigate the more general
situation where the potential is not restricted to inverse square
potential and .One result is that, apart from the inverse square
potential, there are a large number of potentials which can give
the scaling and dominant solution when the function
$\Gamma(\lambda)$ equals $3/2$ for one or some values of
$\lambda_{*}$ as well as the parameter $\lambda_{*}$ satisfies
condition Eq.(18) or Eq.(19). We also find that for a class of
different potentials the dynamics evolution of the universe are
actually the same and therefor undistinguishable.
\\
 {\hspace*{15pt}\small \\ {\bf Keywords:} Scaling Solution; Dark Energy; three-dimensional autonomous
dynamical system; Tachyon Scalar Field.}\\ {\bf PACS:} 98.80.-k,
95.36.+x}
\end{minipage}
\end{center}

\newpage

\section{Introduction} The tachyon field was first arising in the context of string theory \cite{1, 2} and then had been proposed in
cosmology to drive the early inflation \cite{3}-\cite{18}. After
the important finding of an accelerating expansion of our
universe, it became one of the candidates of dark energy to bring
on the late time accelerating expansion\cite{19}-\cite{a1}.
Motivated by the possibility that the equation of state may be
less than $-1$, its phantom version had also been
investigated\cite{31, 32}. Moreover, the cosmological evolution of
its quintom version had been considered in literature\cite{33}.
For this tachyon-quintom model, the dark energy is composed of the
tachyon scalar field as well as the phantom tachyon. These two
scalar fields enable the equation of state $w$ to change from
$w>-1$ to $w<-1$, just like the quintom model of canonical scalar
field. The dynamical evolution of FRW universe filled with a
tachyonic fluid plus a barotropic fluid has been extensively
studied by performing the phase-plane analysis\cite{27}. However,
The potentials of most of the paper about tachyon scalar field are
chosen as the inverse square form\cite{19, 20, 25, 27, 28, 29, 32,
42} partly because that only the inverse potential allows
constructing a two dimensional autonomous dynamical system using
the evolution equations, whereas for any other potentials the
number of dimensions will be higher if the system is to remain
autonomous\cite{27}. The role of inverse square potential in
tachyon scalar field is very similar with the exponential
potential in canonical scalar field(quintessence) model\cite{36,
37, 38}, where only the exponential potential gives a two
dimensional autonomous dynamical system. For the more complicated
case that $\lambda$ is a dynamically changing quantity, i.e. the
potential is not the inverse square form, authors had classified
any type of tachyon potentials to three classifications and
investigated their asymptotic dynamical behavior when $\lambda$
asymptotically approaches to $0$ and $\pm \infty$\cite{28}. They
applied the discussion of constant $\lambda$ to the case of
varying $\lambda$ and obtained the so-called "instantaneous"
critical points. For example, if $\Gamma$ equals the constant
$\Gamma=(n+1)/n$($0<n <2$), the corresponding potential is the
inverse power law potential $V(\phi)=V_0{\phi}^{-n}$. The critical
point $P_5$(see the Table 1) will approach to the dark energy
scenario in which the universe exhibits an accelerating expansion
at late times and the universe in the future will be dominated by
the tachyon scalar field since $x(N)\approx
\lambda(N)/\sqrt{3}\rightarrow 0$ and $y(N)\approx
1-\lambda(N)^2/12 \rightarrow 1$ as $\lambda(N)\rightarrow
0$\cite{28}. Though their classification provided a very useful
way to investigate the cosmological evolution for any type of
tachyon potentials, their method is numerical and the
"instantaneous" critical point is not a true critical point. Here
we will provide an exact method to research a large number of
potentials beyond the inverse square potentials. This method had
been used to investigate the cosmological evolution of the
quintessence model with many different potentials as well as the
well-known exponential potentials\cite{38} and it had been
forwarded to other model\cite{41, 43, 44}. Our method is
considering the parameter $\Gamma$ as a function of $\lambda$. In
this case, the dynamical system(Eqs.(\ref{eq7}-\ref{eq9})) becomes
the three-dimensional dynamical autonomous system. Regarding
parameter $\Gamma$ as a function of $\lambda$ helps us investigate
the cosmological evolution with different tachyon potentials
exactly. In principle, the potential can be figured out via the
relation between parameter $\Gamma$ and $\lambda$, so giving a
concrete form of function $\Gamma(\lambda)$ is equivalent to give
a concrete form of potential $V(\phi)$.  What are the general
properties of the critical points when we consider the higher
three-dimension autonomous system? Does there still exist scaling
solution when we consider other potentials beyond the inverse
square potential? Which critical points exist for all tachyon
scalar field models and which are only relative to the concrete
potentials? In this paper, we will try to respond to these issues.
The paper is organized as follows: in Section 2 we present the
theoretical framework and give the differential relation between
the function $\Gamma(\lambda)$ and potential $V(\phi)$. We find
out all the critical points and investigate their properties in
Section 3. The cosmological implications of these critical points
as well as the summary are presented in section 4.
\section{Basic theoretical frame} We start with a spatially flat Friedman-Robertson-Walker universe containing a scalar field $\phi$ and
a barotropic fluid (with state equation $p_b=(\gamma-1)\rho_b$).
For the sake of simplicity and convenience, we present the basic
equations directly:

\begin{equation}p=L(X,\phi)=-V(\phi)\sqrt{1-\epsilon \dot \phi^2} \label{eq1} \end{equation}
\begin{equation}\rho=2L_{X}X-L(X, \phi)=\frac{V(\phi)}{\sqrt{1-\epsilon \dot \phi^2}} \label{eq2} \end{equation}
\begin{equation}H^2=\frac{\kappa^2}{3}[\frac{V(\phi)}{\sqrt{1-\epsilon \dot \phi^2}}+\rho_b] \label{eq3} \end{equation}
\begin{equation}\dot{H}=-\frac{\kappa^2}{2}[ \frac{\epsilon\dot\phi^2 V(\phi)}{\sqrt{1-\epsilon \dot \phi^2}}+\gamma\rho_b] \label{eq4} \end{equation}

Where $X=\dot \phi^2/2, L_X=\partial L(X, \phi)/\partial X$.
$\epsilon =1$ for tachyon and $-1$ for phantom tachyon field.
"$\cdot$" means the derivative with respect to time $t$.
 The motion equation of the scalar field $\phi$ is:

\begin{equation}\ddot{\phi}+3H(1-\epsilon\dot\phi^2)\dot{\phi}+\frac{\epsilon V'}{V}(1-\epsilon\dot\phi^2)=0 \label{eq5} \end{equation}

We defined the dimensionless variables as follows:

\begin{equation}x=\dot\phi, y=\frac{\kappa\sqrt{V}}{\sqrt3 H}, \lambda=\frac{V'}{\kappa V^{3/2}}, \Gamma=\frac{V V''}{V'^2} \label{eq6} \end{equation}

Where $V'=dV(\phi)/d\phi, V''=d^2V(\phi)/d\phi^2 $. With
Eq.~(\ref{eq6}), Eq.~(\ref{eq4}-\ref{eq5}) can be rewritten in the
following dynamical form:

\begin{equation}\frac{dx}{dN}=-\sqrt{3}(1-\epsilon x^2)(\sqrt{3}x+\epsilon \lambda y) \label{eq7} \end{equation}
\begin{equation}\frac{dy}{dN}=\frac{\sqrt{3}}{2}y(\lambda xy+\frac{\sqrt{3} y^2(\epsilon x^2-\gamma)}{\sqrt{1-\epsilon x^2}}+\sqrt{3}\gamma) \label{eq8} \end{equation}
\begin{equation}\frac{d\lambda}{dN}=\sqrt{3} x y\lambda^2(\Gamma-\frac{3}{2})\label{eq9}\end{equation}

where $N=ln(a)$. The constraint equation from Eq.~(\ref{eq3}) is:
\begin{equation}\frac{y^2}{\sqrt{1-\epsilon x^2}}+\frac{\kappa^2 \rho_b}{3H^2}=1 \label{eq10} \end{equation}

 If the potential is inverse square potential,
$\lambda$ is a constant and $\Gamma$ equals $3/2$, then the three
dimensional dynamical system Eqs.(\ref{eq7}-\ref{eq9}) reduces to
the two dimensional dynamical autonomous system. If we consider
the more complicated case that $\lambda$ is a dynamically changing
quantity($\Gamma \neq 3/2$),then Eqs.(\ref{eq7}-\ref{eq9}) is not
a autonomous system any more since the parameter $\Gamma$ is
unknown. In this case, we can not analyze the evolution of
universe like the inverse square potential exactly. In the
paper\cite{28} authors applied the investigation of constant
$\lambda$ to the dynamically changing $\lambda$ and obtained the
so-called "instantaneous" critical points. Here we propose another
method, which can analyze the evolution of the universe exactly
when the potential is not the inverse square potential. Since
$\lambda$ is the function of tachyon field $\phi$ and $\Gamma$ is
also the function of $\phi$, generally speaking, $\Gamma$ can be
expressed as a function of $\lambda$,

\begin{equation}f(\lambda)=\Gamma(\lambda)-\frac{3}{2} \label{eq11} \end{equation}

then Eq.(\ref{eq9}) becomes:

\begin{equation}\frac{d\lambda}{dN}=\sqrt{3} x y\lambda^2f(\lambda) \label{eq12} \end{equation}

Hereafter, Eq.(\ref{eq7}-\ref{eq8}) and Eq.(\ref{eq12}) are
definitely a dynamical autonomous system. We will show you later
that considering $\Gamma$ as a function of $\lambda$ can cover
many potentials. The three-dimension autonomous system reduces to
two-dimension autonomous systems when $f(\lambda)=0$(i.e,
$\Gamma=3/2$ and $\lambda=constant$).
 \par What is the form of the potential if a function of $f(\lambda)$ is given? We
start with

\begin{equation}\frac{d\lambda}{dV}=\frac{d\lambda}{d\phi}\frac{d\phi}{dV}=\frac{1}{V'}\frac{d}{d\phi}(\frac{V'}{\kappa V^{3/2}})=\frac{\lambda}{V}f(\lambda) \label{eq13} \end{equation}

Integrating Eq.({\ref{eq13}}), we can get the exact function
$\lambda(V)$ with respect to $V$. Then inserting the function
$\lambda(V)$ into the definition of $\lambda$(Eq.(\ref{eq6})),  we
obtain following differential equation for potential $V(\phi)$:

\begin{equation}\frac{dV}{V^{3/2}\lambda(V)}=\kappa d \phi  \label{eq14} \end{equation}

Solving Eq.(\ref{eq14}) will give us the expression of potential
$V(\phi)$. If $f(\lambda)=0$, i.e. $\Gamma=\frac{3}{2}$,
Eqs.(\ref{eq13}-\ref{eq14}) will give the potential $V(\phi)$ with
the form of $V(\phi)=(\frac{1}{2}\kappa \lambda \phi-c_1)^{-2}$,
which is the well-known inverse square potential and has been
studied in detail\cite{19, 20, 25, 27, 28, 29, 32, 42}. However,
using the method in this paper, there are a number of potentials
which can be discussed in principle. Some functions of
$f(\lambda)$ and its corresponding potentials $V(\phi)$ are as
follows:

$$ V(\phi)=(\frac{1}{2}\kappa \lambda \phi-c_1)^{-2} \ \ for \ \ \Gamma(\lambda)=\frac{3}{2}$$
$$ V(\phi)=c_2(\phi-c_1)^{\frac{-2}{2n+1}} \ \ (n\neq -\frac{1}{2}) \ \ for \ \ \Gamma(\lambda)=n+\frac{3}{2}$$
$$ V(\phi)=V_0 e^{\alpha \phi} \ \ (n=-\frac{1}{2}) \ \ for \ \ \Gamma(\lambda)=n+\frac{3}{2}$$
$$ V(\phi)=\frac{V_0}{\phi^2-\phi_0^2} \ \ for \ \ \Gamma(\lambda)=2(1-\frac{1}{\kappa^2V_0\lambda^2})$$
$$ \frac{\beta-c_1V(\phi)^{-\alpha}}{\sqrt{V(\phi)}}=\frac{1}{2}\kappa \alpha \phi+c_2 \ \ (\alpha \neq -\frac{1}{2}) \ for \ \ \Gamma(\lambda)=\beta\lambda+(\alpha+\frac{3}{2})$$
$$ 2\beta V(\phi)^{-\frac{1}{2}}+c_1\alpha ln(V(\phi))=\kappa \alpha \phi+c_2 \ \ (\alpha = -\frac{1}{2}) \ for \ \ \Gamma(\lambda)=\beta\lambda+(\alpha+\frac{3}{2})$$
$$ -\frac{2(-\frac{\beta}{\alpha})^{\frac{1}{n}} \ Hypergeom \left([-\frac{1}{n}, \frac{1}{2}\frac{1}{\alpha n}], [1+\frac{1}{2}\frac{1}{\alpha n}],\frac{c_1\alpha V(\phi)^{-\alpha n}}{\beta}\right) } {\kappa \sqrt{V(\phi)}} =\phi+c_2\ for \ \ \Gamma(\lambda)=\beta\lambda^n+(\alpha+\frac{3}{2})$$
\section{Critical Points and their Properties}
\par The critical points can be found by setting $dx/dN=dy/dN=d\lambda/dN=0$ and their properties are determined by the eigenvalues of the Jacobi matrix
of the three-dimension nonlinear autonomous system Eqs.(\ref{eq7},
\ref{eq8},\ref{eq12}). The eigenvalues of each point are obtained
by linearizing this nonlinear system around each point. All the
points we found are listed in table 1.

{ \begin{center}\begin{tabular}{|c|c|c|c|}\hline Table 1 & $(\lambda_c, x_c, y_c)$ & eigenvalues & Stability \\
 \hline  $P_1(\epsilon=1)$&$\lambda_{a}, 0, 0$&$-3, 3\gamma/2, 0$
& saddle point\\ \hline $P_2(\epsilon=1)$&$\lambda_{a}, \pm 1, 0$&
$6,3\gamma/2, 0$& unstable node\\ \hline $^{*}P_3(\epsilon=1)$&$0,
0, \pm 1$&$ -3, -3\gamma, 0$& $*$\\ \hline
$P_4(\epsilon=1)$&$\lambda_{*}, \mp \sqrt{\gamma},
\pm\frac{\sqrt{3\gamma}}{\lambda_{*}}$&$ \frac{3}{4}[(\gamma-2)
\pm \sqrt{\mu}], -3\lambda_{*}\gamma df_{*}$& Eq.(\ref{eq15})
\\ \hline $P_5(\epsilon=1)$&$\lambda_{*},
\mp\frac{\sqrt{3}\lambda_{*}y_s}{3}, \pm y_s
$&$-3+\frac{\lambda_{*}^2 y_s^2}{2}, -3\gamma+\lambda_{*}^2 y_s^2,
-\lambda_{*}^2 y_s^2 \lambda_{*} df_{*}$ & Eq.(\ref{eq16})
\\ \hline $P_6(\epsilon=-1)$&$\lambda_{a}, 0, 0$&$-3, 3\gamma/2, 0$ & saddle point\\
\hline $^{*}P_7(\epsilon=-1)$&$0, 0, \pm 1$&$-3, -3\gamma, 0$ & $*$\\
\hline $P_8(\epsilon=-1)$&$\lambda_{*},
\pm\frac{\sqrt{3}\lambda_{*}y_c}{3}, \pm y_c $&
$-3-\frac{\lambda_{*}^2 y_c^2}{2}, -3\gamma-\lambda_{*}^2 y_c^2,
\lambda_{*}^2 y_c^2 \lambda_{*}df_{*}$ & Eq.(\ref{eq17})
\\ \hline \end{tabular}\end{center}}
 \footnotetext{$^{*}$ Here one of the eigenvalues of point $P_3$ and $P_7$ is zero and the
rest eigenvalues are negative, this point is called nonhyperbolic
point. the stability of this point cannot be simply determined by
the linearization method and need to resort to other method, for
instance, the center manifold theorem\cite{39}. The center
manifold theorem can help us find the sufficient conditions of
stability of the critical systems but it is somehow
complicated\cite{38}. Another method is to  calculate the three
dimensional nonlinear system Eqs.(\ref{eq7},\ref{eq8},\ref{eq12})
directly and then plot the phase plane to find the critical
point's property numerically.}
 where $\lambda_a$ means an arbitrary value of $\lambda$ and
$\lambda_{*}$ is the value which makes the function
$f(\lambda_{*})=0$, $df_{*} \equiv
\frac{df(\lambda)}{d\lambda}|_{\lambda_{*}}$.

\begin{equation}\mu=17\gamma^2-20\gamma+4+48\gamma^2\sqrt{1-\gamma}/\lambda_{*}^2\label{mu1}\end{equation}

\begin{equation}y_s=\sqrt{\frac{\sqrt{\lambda_{*}^4+36}-\lambda_{*}^2}{6}}, \ \ y_c=\sqrt{\frac{\sqrt{\lambda_{*}^4+36}+\lambda_{*}^2}{6}}\label{ysyc}\end{equation}

\par For the point $P_4$, we have $0<\gamma<1$ since
$\Omega_{\phi}=\frac{3\gamma}{\lambda_{*}^2\sqrt{1-\gamma}}$. We
can also get the condition $
\gamma<\frac{\lambda_{*}^2}{18}(\sqrt{\lambda_{*}^4+36}-\lambda_{*}^2)$
from $\Omega_{\phi}<1$. In fact,  Eq.(\ref{mu1}) can be rewritten
as follows:

\begin{equation}\mu=(\gamma-2)^2-16\gamma(1-\gamma)(1-\Omega_{\phi}),\label{mu2}\end{equation}
and then the real part of its eigenvalues $\frac{3}{4}[(\gamma-2)
\pm \sqrt{\mu}]$ are always negative, so the point $P_4$ will be a
stable node or stable spiral(dependent on the sign of $\mu$) when
the condition is satisfied below:
\begin{equation}\gamma< \frac{\lambda_{*}^2}{18}(\sqrt{\lambda_{*}^4+36}-\lambda_{*}^2) \ and \  \lambda_{*} df_{*}>0 \label{eq15} \end{equation}

\par For the eigenvalues of $P_5$, $\mu_1=-3+\frac{\lambda_{*}^2
y_s^2}{2}=-\frac{3\sqrt{\lambda_{*}^4+36}}{\sqrt{\lambda_{*}^4+36}+\lambda_{*}^2}<0$,
$\mu_2=-3\gamma+\lambda_{*}^2
y_s^2=3[\frac{\lambda_{*}^2}{18}(\sqrt{\lambda_{*}^4+36}-\lambda_{*}^2)-\gamma].$
So $P_5$ is a stable node when:

\begin{equation}\gamma> \frac{\lambda_{*}^2}{18}(\sqrt{\lambda_{*}^4+36}-\lambda_{*}^2) \ and \  \lambda_{*} df_{*}>0 \label{eq16} \end{equation}

Eq.(\ref{eq15}) and Eq.(\ref{eq16}) tell us that $P_4$ and $P_5$
can never be stable in the same time.
\par For the eigenvalues of $P_8$, $\mu_1=-3-\frac{\lambda_{*}^2
y_s^2}{2}<0$, $\mu_2=-3\gamma-\lambda_{*}^2 y_s^2<0$, So $P_8$ is
a stable node when:

\begin{equation} \lambda_{*} df_{*}<0 \label{eq17} \end{equation}

\par The density parameter of tachyon field $\Omega_{\phi}$, the equation of state $w_{\phi}$ and the decelerating factor $q$ are:

\begin{equation}\Omega_{\phi}=\frac{y^2}{\sqrt{1-\epsilon x^2}} \label{eq} \end{equation}
\begin{equation}\gamma_{\phi}=1+w_{\phi}=\epsilon x^2 \label{eq} \end{equation}
\begin{equation}q=\frac{3}{2}[(\gamma-\frac{2}{3})+(\gamma_{\phi}-\gamma)\Omega_{\phi}] \label{eq} \end{equation}

Other properties of these critical points are listed in table 2.
{{\begin{center}\begin{tabular}{|c|c|c|c|c|c|} \hline
Table 2 &$(\lambda_c,x_c, y_c)$&$\gamma_{\phi}$& $\Omega_{\phi}$ & q\\
\hline $P_1$&$\lambda_{a}, 0, 0$    &  0  & 0& $(3\gamma-2)/2 $ \\
\hline $P_2$&$\lambda_{a}, \pm 1, 0$&  1  & Undefined & --  \\
\hline $^{*}P_3$&$0, 0, \pm 1$& 0 & 1 & -1 \\
\hline $P_4$&$\lambda_{*}, \mp \sqrt{\gamma}, \pm\frac{\sqrt{3\gamma}}{\lambda_{*}}$&$\gamma$ & $\frac{3\gamma}{\lambda_{*}^2\sqrt{1-\gamma}}$ & $(3\gamma-2)/2 $   \\
\hline $P_5$&$\lambda_{*}, \mp\frac{\sqrt{3}\lambda_{*}y_s}{3}, \pm y_s $& $\frac{\lambda_{*}^2}{18}[\sqrt{\lambda_{*}^4+36}-\lambda_{*}^2]$ & 1 & $(3\gamma_{\phi}-2)/2 $   \\
\hline $P_6$&$\lambda_{a}, 0, 0$& 0 & 0 & $(3\gamma-2)/2 $   \\
\hline $^{*}P_7$&$0, 0, \pm 1$& 0 & 1 & -1 \\
\hline $P_8$&$\lambda_{*}, \pm\frac{\sqrt{3}\lambda_{*}y_c}{3}, \pm y_c $&-$\frac{\lambda_{*}^2}{18}[\sqrt{\lambda_{*}^4+36}+\lambda_{*}^2]$& $1$& $(3\gamma_{\phi}-2)/2 $   \\
\hline

\end{tabular}\end{center}}

\section{Cosmological Implications}
\hspace*{16 pt} The advantage to investigate the three-dimensional
dynamical system is that we can understand the dynamical evolution
of universe more deeply, though the process will be more
complicated. We find some new critical points which have not been
found previously in the two-dimensional system. Another important
result is that, besides the inverse square potential, there are
many other potentials which also admit the scaling and dominant
solutions. Moreover, from the point of view of three-dimensional
system, we can find out which critical points exist for all
tachyon field and which are only relative to the concrete
potentials.
\par {\it Tachyon Field($\epsilon=1$)}: Points $P_{1-5}$ are the whole
critical points of tachyon field. Here we do not intend to repeat
their properties one by one since the dynamics of tachyon field
has been investigated in detail in literatures\cite{27, 28, 40,
41}. However, from the viewpoint of three dimension, we can get
some new conclusions. Of all the points, points $P_{1-3}$ are
independent of the function $f(\lambda)$ and therefore are nothing
to do with the potential $V(\phi)$ while Points $P_{4-5}$ are
dependent of the concrete potentials. Points $P_4$ and $P_5$,
responding to the scaling and dominant solution, even exist only
when the function $f(\lambda)$ can be zero for one or more values
of $\lambda_{*}$. For example, there are no $P_{4-5}$ for
potential $V(\phi)=c_2(\phi-c_1)^{\frac{-2}{2n+1}} \ (n\neq 0).$
By analyzing three dimension dynamical system, we can study many
potentials and get the detailed dynamical evolution of universe.
However, our results also show that the properties of most
critical points, such as the density parameter $\Omega_{\phi}$,
the decelerating factor $q$, are the same for different potentials
if they satisfy some conditions. That means the dynamics evolution
of universe for a class of different potentials are
undistinguishable. Maybe what we need to do is to research a class
of potentials, not just one special potential. Reader maybe has
found that most critical points and their properties in table 1-2
are the same with inverse square potential\cite{27, 28, 40, 41}.
Another result we should emphasize is the point $P_3$, which
corresponds to the state that universe is dominated by the dark
energy with its equation of state $w_{\phi}$ being $-1$. Moreover,
$P_3$ is a new critical point which does not exist in two
dimension system(namely, when potential being inverse square
potential). In fact, we find that this critical point is just the
"instantaneous" critical point investigated in\cite{28,40}.

\par {\it Phantom Tachyon Field($\epsilon=-1$)}: When $\epsilon$ equals $-1$,
Eqs.(\ref{eq7}-\ref{eq9}) describe the phantom tachyon field with
the kinetic term being negative. This is a quite crazy idea but
has not been excluded by observations. There are only three
critical points($P_{6-8}$) for phantom tachyon field. It is quite
interesting that all the properties of the point $P_6$ is the same
as point $P_1$. Both of them are the saddle points and correspond
to the state that our universe is dominated by the barotropic
fluid. This fact maybe indicates that, no matter the dark energy
is phantom or non-phantom, our universe had truly experienced a
stage that dominated by barotropic matter. The properties of $P_7$
is also the same as $P_3$. For point $P_7$, the tachyon field
behaves as cosmological constant($\gamma_{\phi}=0$) and the
universe is dominated by dark energy($\Omega_{\phi}=1$). As we
have pointed out before, they are the nonhyperbolic points and
their stability cannot be simply determined by the linearization
method. Point $P_8$ is also a stable solution that phantom dark
energy dominated the universe with the equation of state
$\gamma_{\phi}<0$ and density parameter $\Omega_{\phi}=1$. In this
case, the universe will evolve to the "big rip" state inevitably.

\par In summary, in this paper we discuss the three dimensional dynamical
autonomous system of tachyon scalar field directly by taking the
potential parameter $\Gamma$ as the function of another potential
parameter $\lambda$. This is a quite effective method and can be
used to investigate the cosmological evolution as long as the
parameter $\Gamma$ can be expressed as the explicit function of
$\lambda$. From this point of view, the well-known inverse square
potential is just a very easy and special case, a large number of
potentials can be investigated  by this method. When the potential
is inverse square potential, the parameter $\Gamma$ equals $3/2$
and three dimensional autonomous system reduces to the two
dimensional autonomous system. We find an important result that,
besides the extensively discussed inverse square potential, there
are many potentials which can give the tracking solution as long
as the function $\Gamma(\lambda)-3/2=0$ for one or several values
of $\lambda$. Each critical point corresponds to a possible
cosmological state of our universe and its stability tells us how
our universe evolves to this solution. We find that the existence
and properties of some critical points are independent of concrete
potentials, so the cosmological solution related to those points
are possessed by all tachyon field. For the rest critical points,
their existence and properties are related to concrete potentials.
In addition, We investigate the phantom tachyon field and find a
stable solution corresponding to the equation of state
$w_{\phi}<-1$.
\section{Acknowledgement}
\par W.Fang would like to thank Dr.Zhu Chen for useful advices. This work is partly supported by
National Nature Science Foundation of China under Grant
No.10947146, Shanghai Municipal Science and Technology Commission
under Grant No.07dz22020, and the grant from the Shanghai
Education Commission and Shanghai Normal University.

\end{document}